\documentclass{jps-cp}
\usepackage{txfonts} 

\usepackage{here}
\usepackage{pifont}
\usepackage{wrapfig}
\usepackage{url}
\usepackage{amsmath}
\usepackage{comment}
\usepackage{braket}
\usepackage[right]{lineno} 

\title{Measurement of electrons from charm and beauty hadron decays in p-Pb collisions at $\sqrt{s_{\mathrm{NN}}} = 8.16\,$TeV with ALICE}

\author{Daichi Kawana$^{1}$ for the ALICE Collaboration}

\inst{$^{1}$ University of Tsukuba, 1-1-1, Tennodai, Tsukuba, Ibaraki 305-8577, Japan}

\email{s1720250@s.tsukuba.ac.jp}

\recdate{January 11, 2019}

\abst{
	In this proceedings, we report new measurement of electrons from charm and beauty hadron decays in p-Pb collisions at $\sqrt{s_{\mathrm{NN}}} = 8.16\,$TeV as a function of transverse momentum ($p_{\rm{T}}$).
	The nuclear modification factor and cross section ratio with different collision energies are measured.
	Both results imply cold nuclear matter effects on heavy-flavour production in p-Pb collisions are small enough to be negligible and there is no significant collision energy dependence between $\sqrt{s_{\mathrm{NN}}} = 8.16\,$TeV and  $\sqrt{s_{\mathrm{NN}}} = 5.02\,$TeV.
	}

\kword{Heavy-ion collisions, Quark-Gluon Plasma, Heavy-flavour, p-Pb collisions}

\begin{document}
\maketitle


\section{Introduction}
Due to the larger masses ($m_{\mathrm{c,b}} \gg \Lambda_{\mathrm{QCD}}$), 
heavy quarks (charm and beauty) are produced in hard scatterings processes at early stage of the collisions.
Therefore, charm and beauty are powerful probes to test perturbative QCD calculations in pp collisions and study the properties of the Quark-Gluon Plasma (QGP) produced in Pb-Pb collisions.
Strong suppression of the production of electrons from charm and beauty hadron decays at high $p_{\rm{T}}$ has been observed in Pb-Pb collisions with ALICE at the LHC\cite{HFEPbPb2.76}.
For further understanding of the properties of the QGP such as energy loss mechanism in the dense matter, it is very important to study Cold Nuclear Matter (CNM) effects in order to distinguish the role of  initial- and final-state effects on heavy-flavour production.
CNM effects on heavy quark production can be studied in proton-nucleus collisions.
In this contribution, we present measurement of the production of electrons from charm and beauty hadron decays in p-Pb collisions at $\sqrt{s_{\mathrm{NN}}} = 8.16\,$TeV.
We report the $p_{\rm{T}}$ dependence of the nuclear modification factor ($R_{\rm{pPb}}$) and ratio of cross section  of heavy-flavour decays electrons in p-Pb collisions with different collision energies, $\sqrt{s_{\mathrm{NN}}} = 8.16\,$TeV and $\sqrt{s_{\mathrm{NN}}} = 5.02\,$TeV.

\section{Experimental apparatus and data analysis}
The ALICE detector\cite{alice} is a 10,000\,ton integrated apparatus 26\,m long, 16\,m high and 16 \,m wide.
Particles produced in collisions are reconstructed with excellent particle identification capabilities.
Heavy-flavour hadrons are measured by direct reconstruction of D mesons and charm baryons, or indirectly tagged via single electrons or muons from semi-leptonic decay channels of heavy quarks.
This analysis has been performed using experimental data from p-Pb collisions at $\sqrt{s_{\mathrm{NN}}} = 8.16\,$TeV collected in 2016 by the ALICE detector.
\par
Minimum bias data is triggered by coincidence of two V0 detectors.
The V0 detector consists of two scintillator arrays placed in $2.8 < \eta <  5.1$ and $-3.7 < \eta < -1.7$, covering full azimuth.
For extending $p_{\mathrm{T}}$ reach of electrons, events triggered by the Electro-Magnetic Calorimeter (EMCal) are also analyzed. The EMCal trigger requires clusters on the EMCal over threshold energies (5.5\,GeV and 8\,GeV). 
Tracks of charged particles produced in collisions are reconstructed by the Inner Tracking System (ITS) and the Time Projection Chamber (TPC). 
\par
\begin{figure}[H]
  \begin{minipage}{0.5\hsize}
    \begin{center}
      \includegraphics[width=6.7cm]{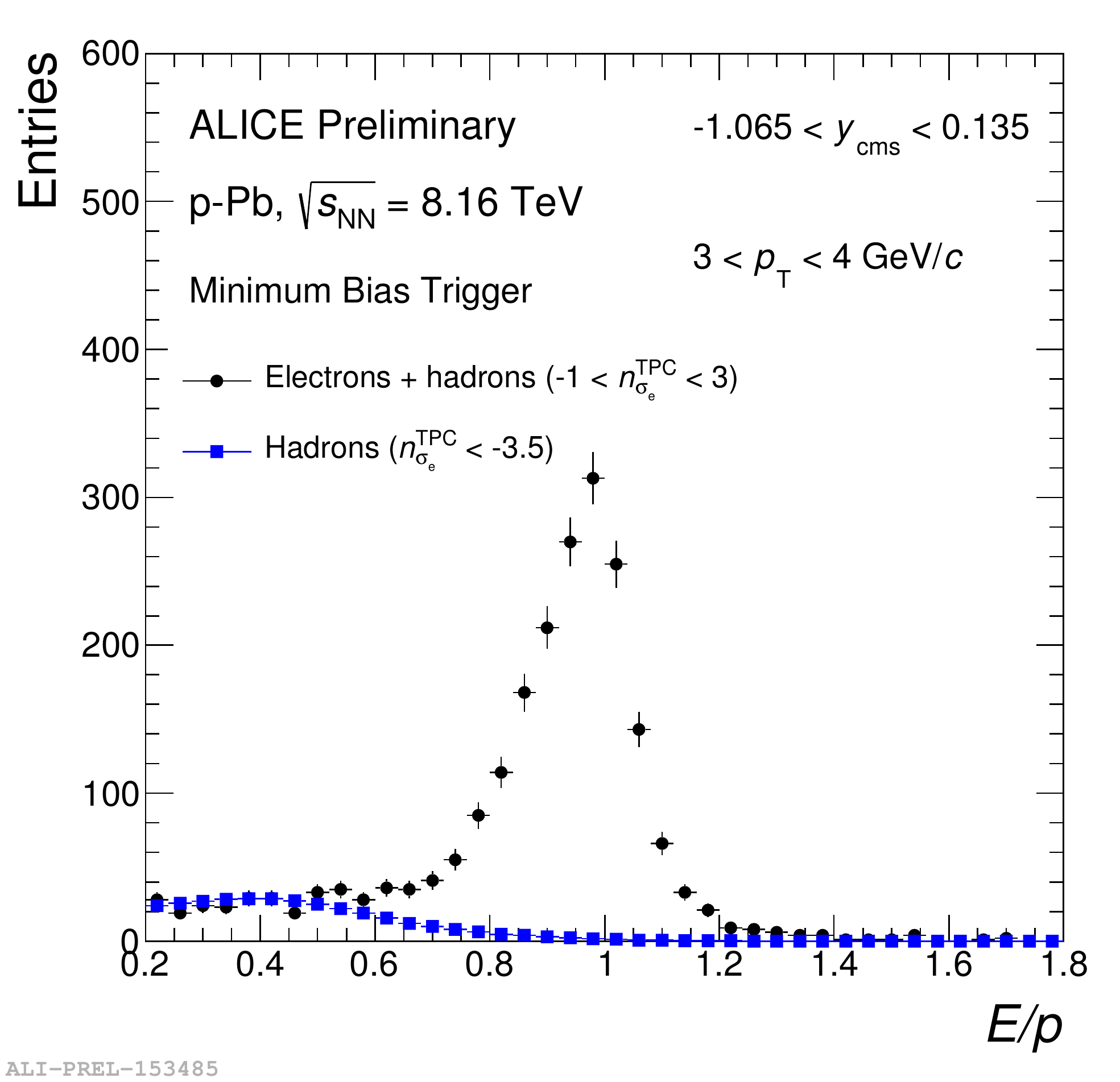}
      \caption{$E/p$ distribution in $3 < p_{\rm{T}} < 4$\,GeV$/c$ in minimum bias trigger events.}
      \label{fig:eop}
    \end{center}
  \end{minipage}
  \begin{minipage}{0.5\hsize}
    \begin{center}
      \includegraphics[width=6.7cm]{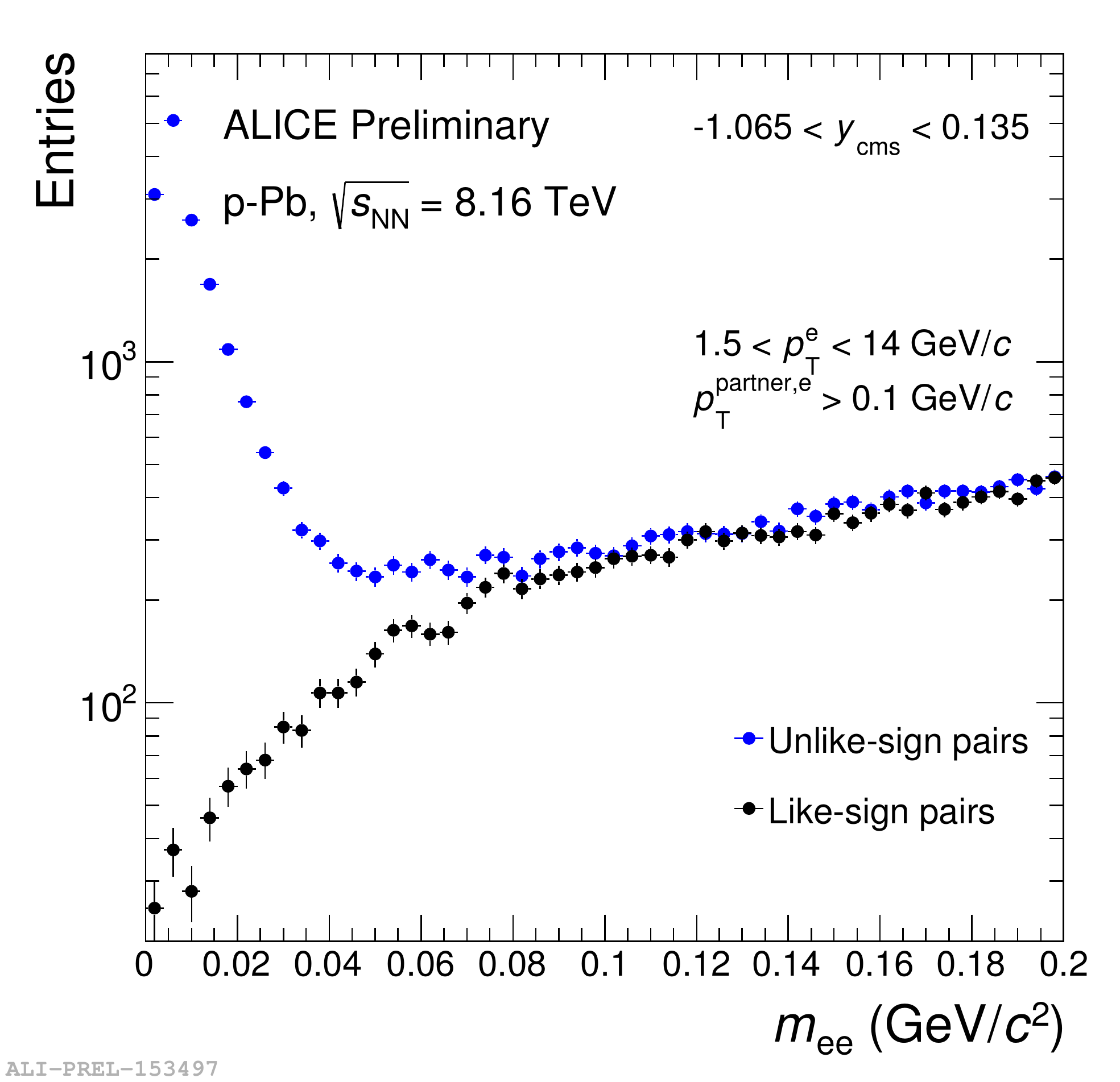}
      \caption{Invariant mass distribution of electron pairs in $1.5 < p_{\rm{T}} < 14$\,GeV$/c$.}
      \label{fig:mass}
    \end{center}
  \end{minipage}
\end{figure}

Electrons can be identified using different detectors in ALICE in different momentum ranges.
In this analysis, electron identification is performed by using the TPC and the EMCal.
Electrons identification with the EMCal is based on the feature of electron; the energy deposit of electrons in the EMCal ($E$) is equal to momentum measured in the TPC ($p$) due to small mass of electrons ($p \gg m_{e}$).
Particles in the TPC are characterized using the ionization energy loss (d$E$/d$x$).
A cut on the shower shape properties in EMCal is also applied to reduce the contamination of hadronic showers.
$E/p$ ratio of electron candidates is shown in Fig. \ref{fig:eop}. and clear peak is observed around $E/p \sim\,1$.
Hadron contamination is estimated with data-driven approach and statistically subtracted from electron candidates. 
$E/p$ of hadron is measured by selecting hadron tracks based on d$E$/d$x$ and scaled to match the $E/p$ between 0.4 and 0.5 where hadrons are dominant in the distribution.

Inclusive electrons contain not only the contribution from charm and beauty hadron decays but also from other decay channels.
Dominant sources of background in $p_{\rm{T}}$ coverage of this analysis are;
\begin{enumerate}
  \item[(a)] electrons from photon conversion in detector material,
  \item[(b)] electrons from Dalitz decays of light neutral mesons (mainly $\pi^0$ and $\eta$),
  \item[(c)] electrons from W-boson decays.
\end{enumerate}
Electrons in (a) and (b) are called photonic electrons, and they can be reconstructed using invariant mass technique.
This approach is based on the feature of photonic electrons; invariant mass of $e^-$ and $e^+$ pairs (Unlike-sign pairs) of photonic electrons candidates is close to zero.
Fig. \ref{fig:mass} shows invariant mass distribution of electron pairs. Signal of photonic electrons in invariant mass distribution is clearly shown around $m_{ee}\sim\,0$ for Unlike-sign pairs. Contribution of combinatorial background is estimated by calculating invariant mass with $e^+\,e^+$ and $e^-\,e^-$ pairs (Like-sign pairs), and statistically subtracted from Unlike-sign pairs. 
Contribution of electrons from W-boson decays are estimated by POWHEG\cite{powheg} simulation.
Finally those background electrons are statistically subtracted from inclusive electrons, then yield of electrons from charm and beauty hadron decays are obtained.
\par
The nuclear modification factor $R_{\rm{pPb}}$ is defined as the equation below:
\begin{equation}
R_{\mathrm{pPb}} = \frac{\mathrm{d} \sigma_{\mathrm{pPb}} / \mathrm{d} p_{\mathrm{T}}}{A \times \mathrm{d} \sigma_{\mathrm{pp}} / \mathrm{d} p_{\mathrm{T}}}.
\end{equation}
where $\mathrm{d} \sigma_{\mathrm{pp}} / \mathrm{d} p_{\mathrm{T}}$ and $\mathrm{d} \sigma_{\mathrm{pPb}} / \mathrm{d} p_{\mathrm{T}}$ are respectively the invariant cross section in pp and p-Pb collisions, and $A = 208$ is number of nucleons in lead nucleus.
To obtain $\mathrm{d} \sigma_{\mathrm{pp}} / \mathrm{d} p_{\mathrm{T}}$ of electrons from charm and beauty hadron decays at $\sqrt{s} = 8\,$TeV, $\sqrt{s}$-interpolation is performed based on FONLL\cite{fonll} and experimental results at $\sqrt{s} = 7\,$TeV are scaled to $\sqrt{s} = 8\,$TeV.
	$\mathrm{d} \sigma_{\mathrm{pp}} / \mathrm{d} p_{\mathrm{T}}$ measured with ALICE\cite{bib:ALICEpp7TeV} is used for  $p_{\rm{T}} < 8$\,GeV/$c$, and that one measured with ATLAS\cite{bib:ATLASpp} is used for $p_{\rm{T}} > 8$\,GeV/$c$.
	
\section{Results and discussions}
\begin{figure}[H]
  \begin{minipage}{0.5\hsize}
    \begin{center}
      \includegraphics[width=6.7cm]{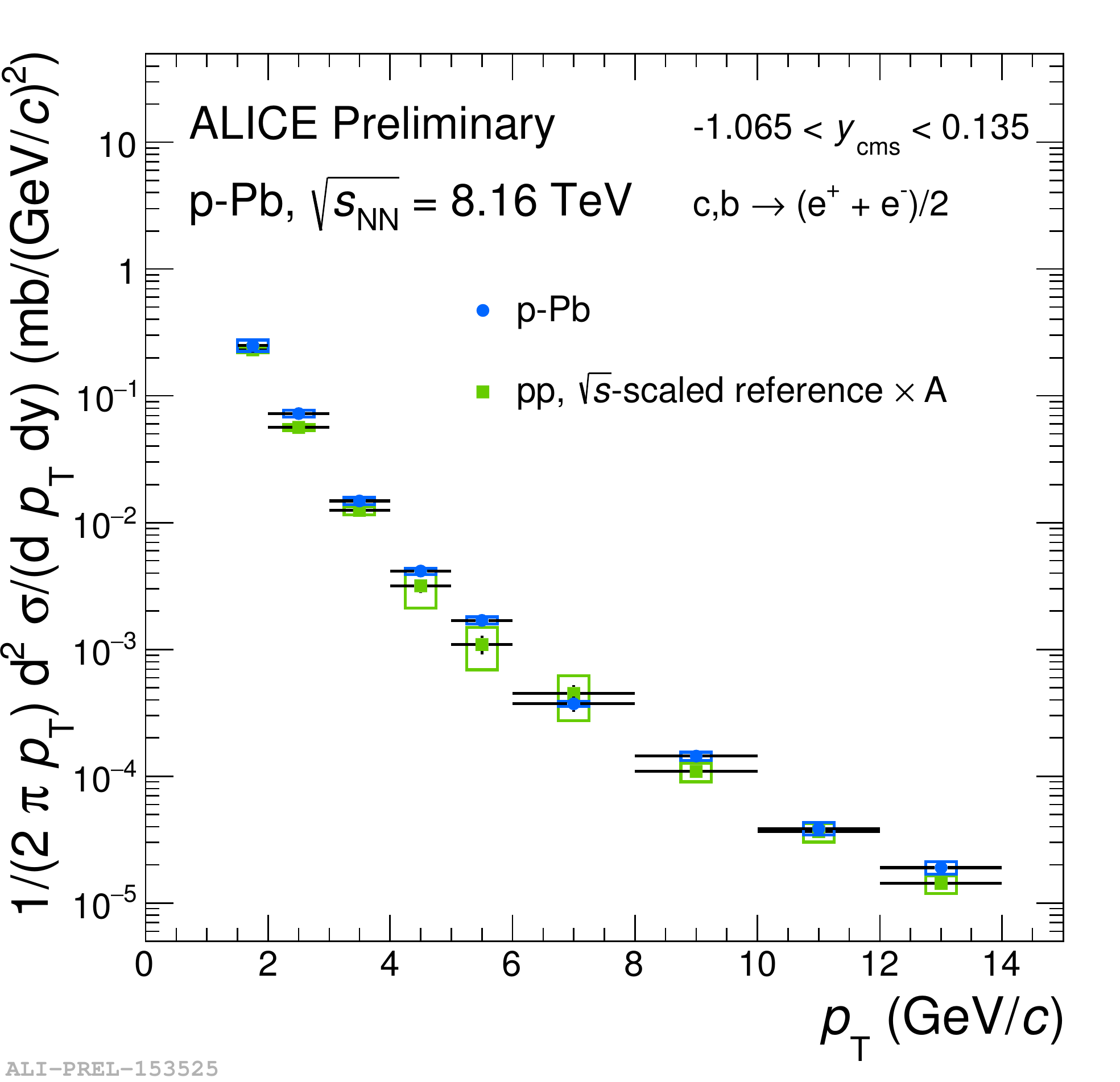}
      \caption{The $p_{\rm{T}}$-differential invariant cross section of electrons from charm and beauty hadron decays in p-Pb collisions at $\sqrt{s_{\mathrm{NN}}} = 8.16\,$TeV and pp reference.}
      \label{fig:cs}
    \end{center}
  \end{minipage}
  \begin{minipage}{0.5\hsize}
    \begin{center}
      \includegraphics[width=6.7cm]{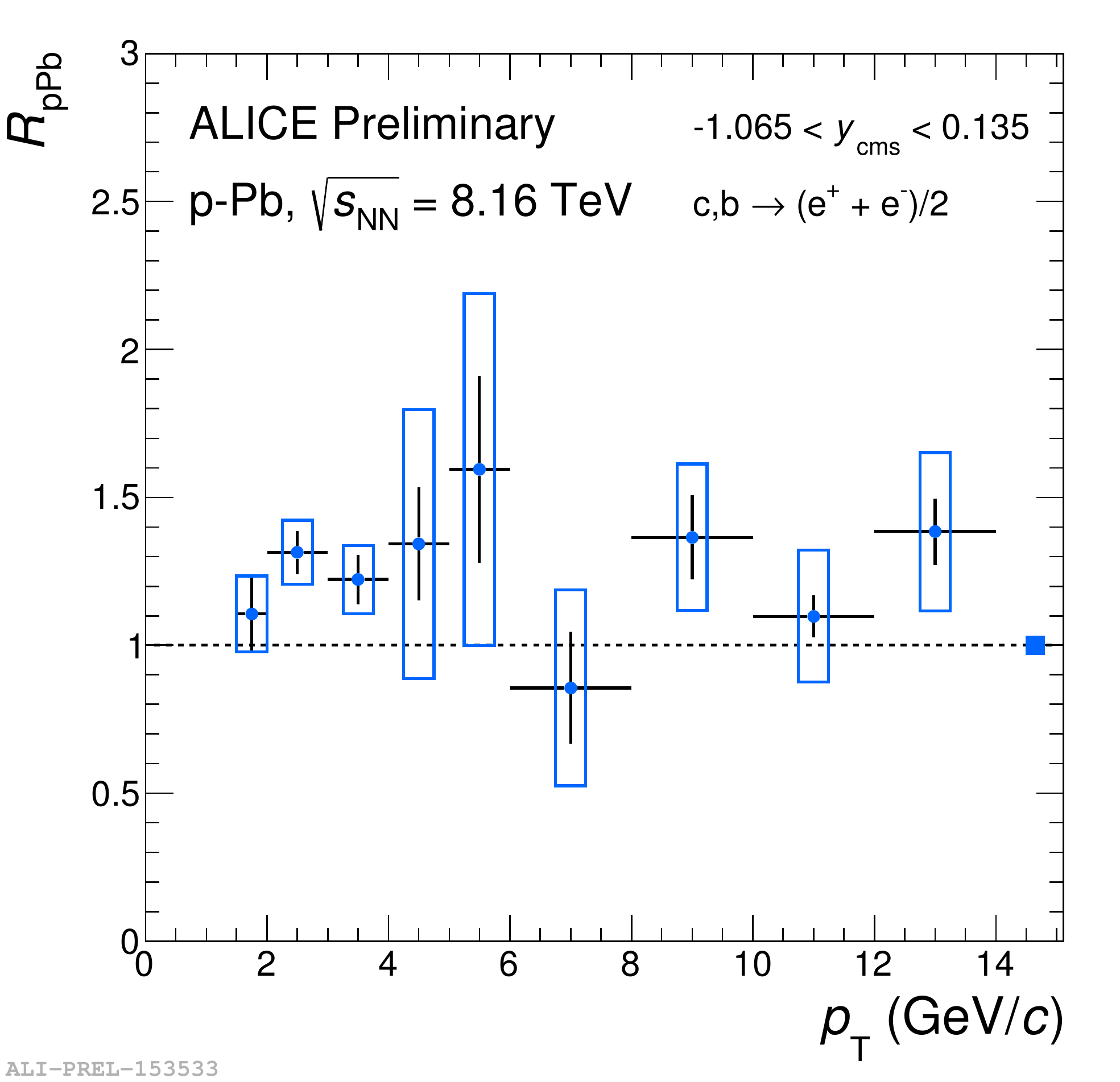}
      \caption{The nuclear modification factor of electrons from charm and beauty hadron decays in p-Pb collisions at $\sqrt{s_{\mathrm{NN}}} = 8.16\,$TeV.}
      \label{fig:RpPb}
    \end{center}
  \end{minipage}
\end{figure}

\begin{figure}[H]
  \begin{minipage}{0.5\hsize}
    \begin{center}
      \includegraphics[width=6.7cm]{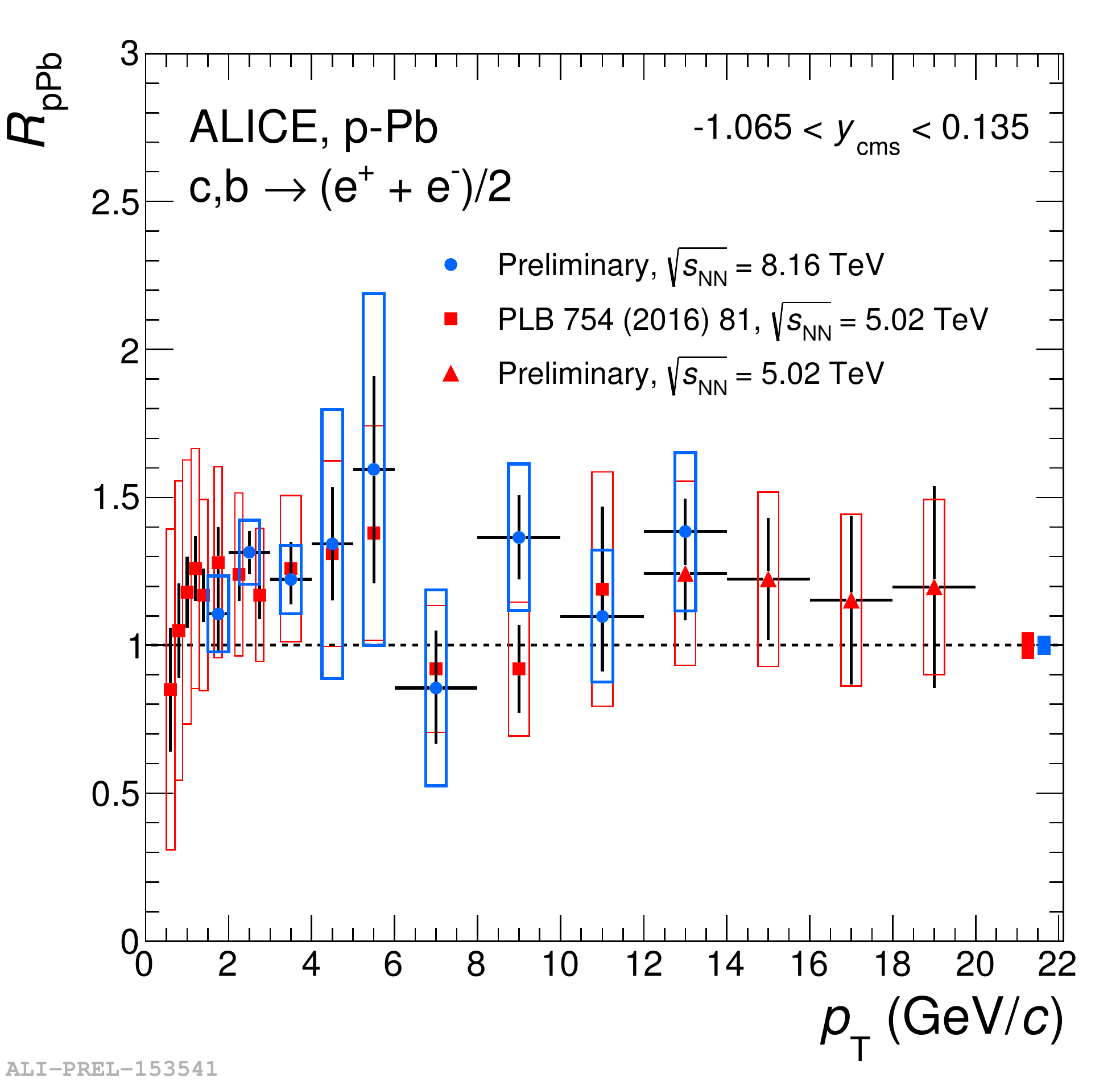}
      \caption{The nuclear modification factor of electrons from charm and beauty hadron decays in p-Pb collisions at $\sqrt{s_{\mathrm{NN}}} = 8.16\,$TeV and $\sqrt{s_{\mathrm{NN}}} = 5.02\,$TeV.}
      \label{fig:RpPbw5TeV}
    \end{center}
  \end{minipage}
  \begin{minipage}{0.5\hsize}
    \begin{center}
      \includegraphics[width=6.7cm]{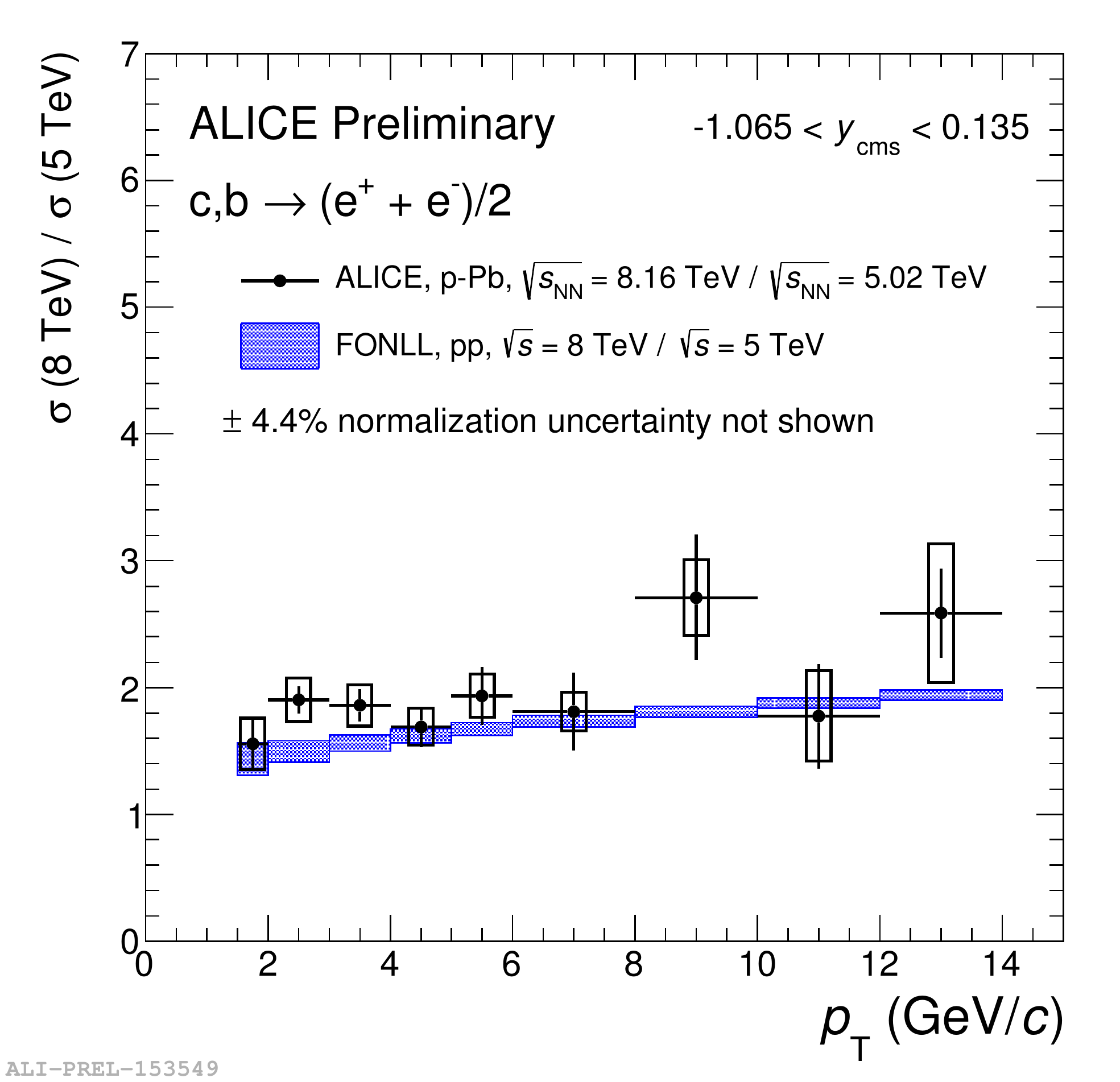}
      \caption{The cross section ratio of electrons from charm and beauty hadron decays in p-Pb collisions at $\sqrt{s_{\mathrm{NN}}} = 8.16\,$TeV and $\sqrt{s_{\mathrm{NN}}} = 5.02\,$TeV.}
      \label{fig:CrossSectionRatio}
    \end{center}
  \end{minipage}
\end{figure}

Fig. \ref{fig:cs}. shows the $p_{\rm{T}}$-differential invariant cross section of electrons from charm and beauty hadron decays in pp and p-Pb collisions. The cross section in pp collisions is scaled by A = 208.
By calculating the ratio of these cross sections, the nuclear modification factor $R_{\mathrm{pPb}}$ can be obtained as shown in Fig. \ref{fig:RpPb}.
The nuclear modification factor is measured in $1.5 < p_{\rm{T}} < 14$\,GeV$/c$, and consistent with unity in the whole $p_{\rm{T}}$ interval within uncertainties. 
Hence heavy-flavour production in p-Pb collisions at $\sqrt{s_{\mathrm{NN}}} = 8.16\,$TeV scales with the number of binary collisions, and the suppression of charm and beauty observed in central Pb-Pb collisions\cite{HFEPbPb2.76} is not present in p-Pb collisions.
In Fig. \ref{fig:RpPbw5TeV} the nuclear modification factor of electrons from charm and beauty hadron decays in p-Pb collisions at $\sqrt{s_{\mathrm{NN}}} = 8.16\,$TeV is compared with the one at $\sqrt{s_{\mathrm{NN}}} = 5.02\,$TeV\cite{bib:ALICEpPb5TeV}, and  a similar trend is observed.
Therefore CNM effects on heavy-flavour production are small/negligible at both energies.
To access directly the $\sqrt{s}$-dependence of CNM effects on heavy-flavour production, ratio of invariant cross section of electrons from charm and beauty hadron decays at the two energies, $\sigma_{\mathrm{pPb}} (\sqrt{s_{\mathrm{NN}}} = 8.16\,\mathrm{TeV})$\,/\,$\sigma_{\mathrm{pPb}} (\sqrt{s_{\mathrm{NN}}} = 5.02\,\mathrm{TeV})$ is calculated and shown in Fig. \ref{fig:CrossSectionRatio}.
The cross section ratio is compatible with the ratio in pp collisions estimated by FONLL calculation shown as blue band in Fig. \ref{fig:CrossSectionRatio}.
Thus this result indicates that there is no significant $\sqrt{s}$-dependence of CNM effects on heavy-flavour production in p-Pb collisions. 

\section{Conclusions}
Electrons from charm and beauty hadron decays in p-Pb collisions at $\sqrt{s_{\mathrm{NN}}} = 8.16\,$TeV are investigated in $1.5 < p_{\rm{T}} < 14$\,GeV$/c$ with the ALICE at the LHC.
The nuclear effects on charm and beauty production in p-Pb collisions are quantified by the nuclear modification factor $R_{\mathrm{pPb}}$, and it is compatible with unity within uncertainties and also with $R_{\mathrm{pPb}}$ at $\sqrt{s_{\mathrm{NN}}} = 5.02 \,$TeV. 
Hence CNM effects on charm and beauty production are small enough to be negligible, and the strong suppression observed in Pb-Pb collisions is due to hot and dense medium effects.
The cross section ratio of heavy-flavour hadron decays electrons in p-Pb collisions $\sigma_{\mathrm{pPb}} (\sqrt{s_{\mathrm{NN}}} = 8.16\,\mathrm{TeV})$\,/\,$\sigma_{\mathrm{pPb}} (\sqrt{s_{\mathrm{NN}}} = 5.02\,\mathrm{TeV})$ is well described by the ratio in pp collisions predicted by pQCD-based theoretical calculation. This result indicates no collision energy dependence of CNM effects on charm and beauty production.

\end{document}